\documentclass[preprint,10pt]{elsarticle}
\usepackage{amssymb}
\usepackage{graphicx}
\usepackage{float}
\usepackage{booktabs}
\usepackage{setspace}
\usepackage{epstopdf}
\usepackage{subfig}
\usepackage[pagewise]{lineno}
\usepackage{array}
\usepackage{natbib}
\newcolumntype{L}[1]{>{\raggedright\let\newline\\\arraybackslash\hspace{0pt}}m{#1}}
\newcolumntype{C}[1]{>{\centering\let\newline\\\arraybackslash\hspace{0pt}}m{#1}}
\newcolumntype{R}[1]{>{\raggedleft\let\newline\\\arraybackslash\hspace{0pt}}m{#1}}
\AppendGraphicsExtensions{.tif}

\biboptions{sort&compress}

\journal{Journal of Sound and Vibration}

\begin{document}

\begin{frontmatter}

\title{Deep learning surrogate interacting Markov chain Monte Carlo based full wave inversion scheme for properties of materials quantification}
\author[label1]{Reza Rashetnia$^{*}$}
\author[label2]{Mohammad Pour-Ghaz}
\address[label1]{$^{*}$InstroTek Inc., 1 Triangle Dr, Research Triangle Park, Durham, NC 27709, USA,
(Corresponding author) Tel.:+1-979-402-0417. E-mail:reza.rashetnia@gmail.com}
\address[label2]{Department of Civil Construction and Environmental Engineering, North Carolina State University, Raleigh, NC 27695, USA.}

\begin{abstract}

Full Wave Inversion (FWI) imaging scheme has many applications in engineering, geoscience and medical sciences.
In this paper, a surrogate deep learning FWI approach is presented to quantify properties of materials using stress waves.
Such inverse problems, in general, are ill-posed and nonconvex, especially in cases where the solutions exhibit shocks, heterogeneity, discontinuities, or large gradients.
The proposed approach is proven efficient to obtain global minima responses in these cases.
This approach is trained based on random sampled set of material properties and sampled trials around local minima, therefore, it requires a forward simulation can handle high heterogeneity, discontinuities and large gradients.
High resolution Kurganov-Tadmor (KT) central finite volume method is used as forward wave propagation operator.
Using the proposed framework, material properties of 2D media are quantified for several different situations.
The results demonstrate the feasibility of the proposed method for estimating mechanical properties of materials with high accuracy using deep learning approaches.

\end{abstract}
\end{frontmatter}
\section{Introduction}

Tomography techniques are valuable and widely used in engineering, geoscience and medical sciences.
Examples of such tomography based techniques include, Infrared Thermography \cite{Bagavathiappan2013, Lahiri2012}, Electrical Impedance Tomography \cite{Rashetnia2018a, Rashetnia2017, Smyl2017, Rashetnia2017a, Brown2009, Davalos2004, Borcea2002}, 
Electrical Capacitance Tomography \cite{Voss2018,Voss2017}, 
radar acoustics and Radiography \cite{Buyukozturk1998, Topczewski2007}, X-Ray Computed Tomography \cite{Balazs2018, Mees2003, Ketcham2001}, and stress wave based tomography \cite{Choi2015, Lin2018, Krause2001, Schickert2005, Haza2013, Beniwal2016, Liu2010, Liu2011, Zhu2007, Yu2019, Beniwal2015, Liu2019, Kawashima2006, Rashetnia2018, Law2005, Lu2006, Rashetnia2016}.
Among these methods, stress wave-based methods are attractive since they can provide information about the mechanical properties of materials and structures.
Stress wave methods are mostly refraction and reflection tomography techniques, which use only the travel time kinematics of the transducer data.
Full Wave Inversion (FWI) is a type of stress wave tomography which uses complete waveforms and derives high resolution velocity models by minimizing the difference between observed and modeled waveforms. 
FWI goes beyond refraction and reflection tomography techniques, which use only the travel time kinematics of the signals, by using additional information provided by the amplitude and phase of the stress waveform.
The highly detailed models provided by FWI can be used to resolve complex mechanical features both in time and frequency domains\cite{Guitton2012,Hu2009,Lin2015a,Lin2015b,Vigh2008}.
There are few research can be applied on deep learning to solve FWI for mechanical behavior reconstruction \cite{Lewis2017,Richardson2018}.
This paper provides a computational inverse framework for stress wave FWI tomography to quantify distribution of elastic moduli and densities at the same time or separately.

For numerical inverse solutions of stress wave tomography, nonlinear least-squares, Newton’s method or other data-fitting methods can be natural choices commonly used for finding the coefficients in the systems of partial differential equations chosen to model the wave physics.
However, they may be cumbersome and computationally expensive to implement for stress wave differential equation specially for finer mesh discretization.
Also, gradient-based methods require proper objective functions and constraints for optimal convergence, and they suffer from inability of optimal convergence to global minima in cases of nonconvexity and high ill-posedness and nonlinearity.
The numerical implementations of FWI is considered highly non-linear, ill-posed and often nonconvex inverse problem \cite{Virieux2009}.
This becomes worst in the case of materials with large properties gradient, nonlinearity or shock waves.
To resolve these issues, we employ a surrogate deep learning interacting Markov chain Monte Carlo based optimization method to solve the FWI inverse problem and estimate properties of materials.
In this paper, a surrogate deep-learning optimization approach is presented to minimize the difference between observed and modeled responses. 
Surrogate deep learning random search naturally causes large properties gradient and high heterogeneity itself.
Therefore, this method requires a wave propagation forward modeling which can handle heterogeneity, large properties gradient, nonlinearity and shock waves properly.

Kurganov-Tadmor (KT) high resolution central finite volume scheme is used for solving stress waves in two dimensional heterogeneous media. 
KT is highly accurate for quantification of materials properties with large properties gradient, nonlinearity or shock waves.
This scheme is non-oscillatory and enjoy the main advantage of Godunov-type central schemes: simplicity, i.e., they employ neither characteristic decomposition nor approximate Riemann solvers. 
This makes it universal method that can be applied to a wide variety of physical problems, including hyperbolic systems of conservation laws. 
KT central scheme has the numerical dissipation with an amplitude of order $O(\Delta X^{3}/\Delta t)$ \cite{Kurganov2000}.
Beside of the good resolution obtained by the KT, it can use a semi-discrete formulation coupled with an appropriate ODE solver retaining simplicity and high resolution with lower numerical viscosity, proportional to the vanishing size of the time step $\Delta t$ \cite{Kurganov2000}.
This semi-discrete central scheme is based on the ideas of Rusanov’s method using a more precise information about the local speeds of wave propagation computed at each Riemann Problem in two-space dimensions \cite{Rusanov1961}.
In this paper, KT is used as forward-wave propagation operator $f$ to map the stress wave velocity to ultrasonic stress wave signals.

In the following, we present KT scheme as the forward model used in this work.
Then, deep learning model used in the inverse problem is described.
This is followed by deep learning surrogate interacting Markov chain Monte Carlo inverse algorithm.
Then, results and discussions are provided and proposed method is investigated on three different examples.
Finally, we summarize our findings in conclusions.
\section{Forward model: Kurganov-Tadmor scheme}
\label{sec.KT}

Kurganov-Tadmor (KT) high resolution central scheme recently developed by Kurganov and Tadmor \cite{Kurganov2000}, which is one of Monotonic Upwind Schemes for Conservation Laws (MUSCL) \cite{Leer1979}.
In order to solve partial differential equations, the MUSCL schemes are finite volume schemes that can provide highly accurate numerical solutions for given systems, even in cases where the solutions exhibit shocks, discontinuities, or large gradients. 
Examples of pioneering works in MUSCL schemes include the first-order Lax Friedrichs scheme \cite{Lax1954} and the Nessyahu-Tadmor (NT) scheme \cite{Nessyahu1990} which offers higher resolution as compared to the former.
Both Lax Friedrichs and NT schemes have a numerical viscosity of the order of $o((\Delta x)^{2r}\setminus\Delta t)$ and suffer from high numerical viscosity when sufficiently small time steps are used $(\Delta t \rightarrow 0)$\cite{Kurganov2000}.
KT has a much smaller numerical viscosity of $(o((\Delta x)^{2r-1}))$ which is independent of the $o(1\setminus\Delta t)$.
KT retains the independence of eigen structure of the problem and approximates the solutions of nonlinear conservation laws and convection-diffusion equations with less effort \cite{Kurganov2000}.
It also requires much lower number of mesh points containing the wave compared with a first-order methods with similar accuracy.
KT provides simple semi-discrete formulation and also uses more precise information of the local propagation speeds which in turn increases the accuracy of the solution.
For these reasons, KT is chosen for this study which let us to achieve more accurate FWI results.
In this study, KT is developed to solve the stress wave propagation,

\begin{equation}
\frac{\partial^{2} u}{\partial t^{2}} = c^{2}\nabla^{2} u
\label{PDE}
\end{equation}
where $u(x,t)$ is the displacement and $c$ is the constant coefficient.

To use the semi-discrete KT scheme, Equation \ref{PDE} need to be expressed in first order hyperbolic equation. 
Therefore, we implement the following change of variables on Equation \ref{PDE}.

\begin{equation}
\frac{\partial}{\partial t} \varepsilon - \frac{\partial}{\partial x} v = 0
\label{1DPDEs_1}
\end{equation}

\begin{equation}
\frac{\partial}{\partial t}(\rho(x) v)-\frac{\partial}{\partial x} \sigma = 0
\label{1DPDEs_2}
\end{equation}
where $v$, $\varepsilon$, $\sigma$, and $\rho$ are velocity, strain, stress, and density respectively. In Equation \ref{1DPDEs_1}, $\varepsilon$ is the state variable, and $v$ is the flux. Similarly, in Equation \ref{1DPDEs_2}, $\rho(x) v$ is the state variable, and $\sigma$ is the flux. 

In KT scheme, linear piecewise approximation of state variable ($\hat{u}$) is shown by Equation \ref{Piecewise} within each cell. Equation \ref{Piecewise} is used to discretize Equations \ref{1DPDEs_1} and \ref{1DPDEs_2} using slope-limited, left and right extrapolated state variable.
Hence, high resolution total variation diminishing discretization can be written as Equation \ref{TVD}

\begin{equation}
\hat{u}(x) = \hat{u}_{i} + \frac{x-x_{i}}{x_{i+1}-x_{i}} (\hat{u}_{i+1}-\hat{u}_{i}) \hspace{3 mm} \forall x \in (x_{i},x_{i+1}]
\label{Piecewise}
\end{equation}

\begin{equation}
\frac{d\hat{u}_{i}}{dt} + \frac{1}{\Delta x_{i}} [f^{*}(\hat{u}_{i+1/2})-f^{*}(\hat{u}_{i-1/2})] = 0
\label{TVD}
\end{equation}
where  $i$ is the cell center index.
The fluxes $f^{*}(\hat{u}_{i\pm1/2})$ are nonlinear combination of the first and second order approximations of the continuous flux function at cell edges. The fluxes, $f^{*}(\hat{u}_{i\pm1/2})$, are calculated based on Equations \ref{Fneg} and \ref{Fpos}.

\begin{equation}
f^{*}(\hat{u}_{i-\frac{1}{2}}) = \frac{1}{2} \{[f(\hat{u}^{R}_{i-\frac{1}{2}})+f(\hat{u}^{L}_{i-\frac{1}{2}})]-a_{i-\frac{1}{2}}[\hat{u}^{R}_{i-\frac{1}{2}}-\hat{u}^{L}_{i-\frac{1}{2}}]\}
\label{Fneg}
\end{equation}

\begin{equation}
f^{*}(\hat{u}_{i+\frac{1}{2}}) = \frac{1}{2} \{[f(\hat{u}^{R}_{i+\frac{1}{2}})+f(\hat{u}^{L}_{i+\frac{1}{2}})]-a_{i+\frac{1}{2}}[\hat{u}^{R}_{i+\frac{1}{2}}-\hat{u}^{L}_{i+\frac{1}{2}}]\}
\label{Fpos}
\end{equation}
where $R$ and $L$ are the right and left cells at the $i\pm\frac{1}{2}$ edges. The local propagation speed in each cell edge, $a_{i\pm\frac{1}{2}}$, is the maximum absolute value of the eigenvalue of the Jacobian of $f(\hat{u}(x,t))$, over cell $i$ and $i\pm1$:

\begin{equation}
a_{i\pm\frac{1}{2}}(t)=max[\textrm{abs}(\rho(\frac{\partial f(\hat{u}_{i}(t))}{\partial \hat{u}})),\textrm{abs}(\rho(\frac{\partial f(\hat{u}_{i\pm 1}(t))}{\partial \hat{u}}))]
\label{F21}
\end{equation}
where $\rho$ is spectral radius of $\frac{\partial f (\hat{u} (t))}{\partial \hat{u}}$.

\begin{figure}[H]
    \centering
    \includegraphics[width=3.5in]{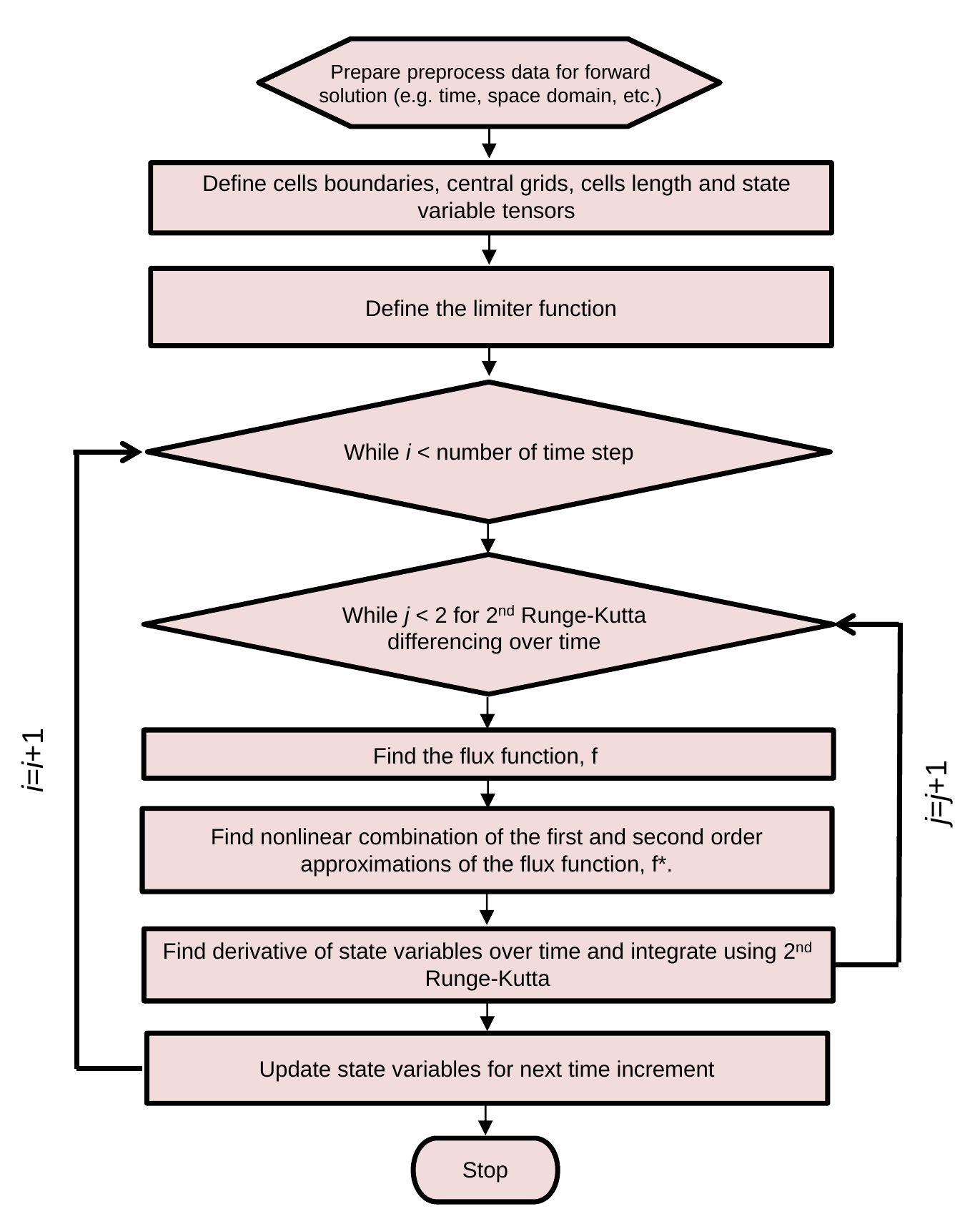}
    \caption{Flowchart of Kurganov-Tadmor central scheme}
    \label{FC1}
\end{figure}

We use $2^{nd}$ Runge Kutta for time integration over $\frac{d\hat{u}_{i}}{dt}$ after it is found by Equation \ref{TVD} \cite{Kurganov2000}.
Figure \ref{FC1} presents the flowchart of KT implementation.

\section{Deep learning model}
\label{DLM}

\begin{figure}[ht!]
    \centering
    \includegraphics[width=3.5in]{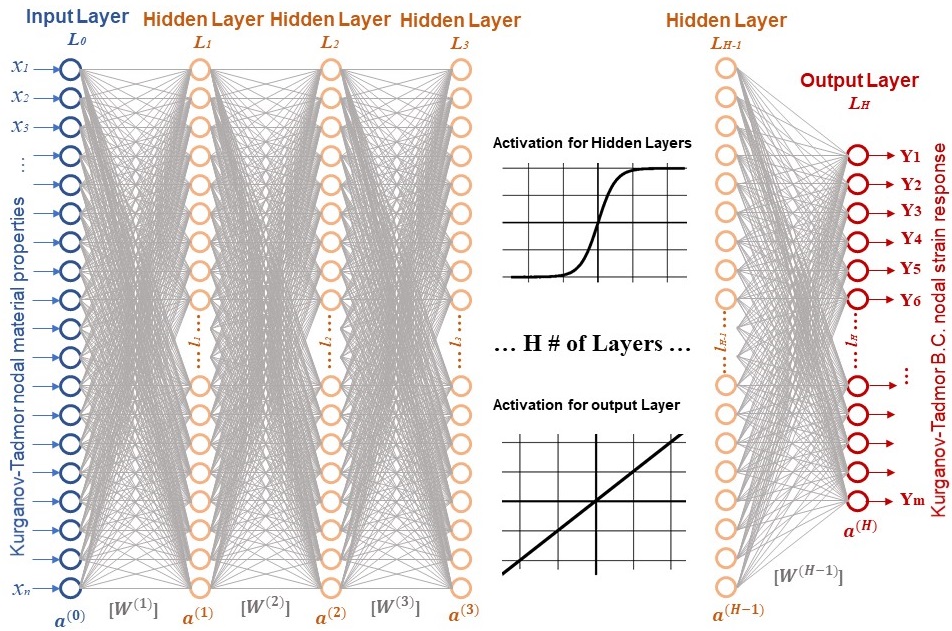}
    \caption{Deep learning architecture.}
    \label{DLM_1}
\end{figure}

In this paper, deep neural network with fully connected layers is used to input nodal elastic moduli and densities and output nodal strains over certain boundaries.
Figure \ref{DLM_1} presents the deep learning architecture in this study.
The architecture of this network includes input layer ($L_{0}$) of nodal KT material characteristics, hidden layers ($L_{1}-L_{H-1}$) and output layer $L_{H}$ which are nodal strain values over certain boundaries.
For these networks, high number of hidden layers are considered to potentially account for the sophisticated nonlinear function mapping the inputs to the outputs, where the number layers and units of these layers will be chosen as hyperparameters based on the network’s performance for each example.
All hidden layers utilize the tanh activation for modelling the nonlinearity, and the output layer has linear activation to reconstruct strain values.
The inverse problem of wave propagation equation is non-convex.
In order to deal with non-convexity of the deep network optimization, the Logistic Regression cost function was used.
Stochastic Gradient Descent (SGD) approach was used for network optimization.
To improve the model performance, $L_{2}$ and dropout regularization approaches were used. 
$L_{2}$ regularization parameters were chosen using an initial batch of random parameter vectors as training and development sets to avoid overfitting (high variance) and underfitting (high bias).
To capture better randomness, dropout method was used.
Dropout probability is chosen for each example as a regularization hyperparameter.
For weight initialization of deep network, Xavier initialization approach is used for faster convergence of optimization.
For faster optimization, also mini-batch gradient descent with momentum and RMSprop (Adam method) is used.
The mini-batch sizes and momentum terms also will be tuned as hyperparameters.
Learning rate and its decay coefficient will be tuned as hyperparameters.
In order to tune all deep-learning hyperparameters such as number of hidden layers, layers' units, regularization parameters, mini-batch sizes, Adam hyperparameters and learning rates, initial deep learning network is trained using KT forward models.

\section{Inverse model: deep learning surrogate interacting Markov chain Monte Carlo search}
 \label{subsect.SMARS}

The inverse problem is addressed as estimation of the mechanical properties of media using stress wave propagation, which has a non-convex error surface with multiple local minima due to data sparsity.
Therefore, the gradient-based optimization algorithms may provide convergence to local minima and computationally become highly expensive. 
Further, high ill-posedness of this inverse problem increases gradient-based algorithms variance which required more complex regularization techniques without guaranteeing better reconstructions. 

In this study, Interacting Markov chain Monte Carlo (IMCMC) algorithm is used as the search algorithm in order to find global minima and handle the data sparsity.
Unlike most of the current MCMC methods that ignore the previous trials; here, optimized deep learning model is used to speed-up the Markov Chain Monte Carlo algorithm by an order of magnitude \cite{Tahmasebi2016}.
In this method, the IMCMC method iteratively samples randomly from a sequence of probability distributions and estimating their errors with respect to waveform measurements which increases level of sampling complexity \cite{DelMoral2013}.
Then, using the deriven data set, a deep learning model is trained based on samplings.
Finally, the deep learning model is optimized to estimate the best unknown parameters in order to provide lowest errors with respect to measurements.
This sequence is iteratively repeated until the global minima is aquired. 
The IMCMC method provides sampling from sequence of probability distributions and avoids convergence to local minima.

The objective is to minimize model error function ($e^{fm}$), per Equation \ref{error}

\begin{equation}
 e^{FM}(\vec{p})=\sum^{m}_{i=1}\|f_{i}^{FM}(\vec{p},\vec{q})-f_{i}^{mes}(\vec{q})\|_{L_{2}(Q)}
\label{error}
\end{equation}
\noindent
where $ \vec{p} \in \mathbb{R}^{n}$ are a set of vectors with $n$ unknown material parameters,
$\vec{q} \in Q$ is the vector of variables in the measurement domain (e.g., time, boundary conditions), $f_{i}^{FM}(\vec{p},\vec{q})$ is the strain response from the forward model (i.e., KT),
$f_{i}^{mes}(\vec{q})$ is the measured strain response, and $m$ is the number of response components.
Similarly, the deep learning model error function is defined per Equation \ref{error2}

\begin{equation}
e^{ML}(\vec{p})=\sum^{m}_{i=1}\|f_{i}^{ML}(\vec{p},\vec{q})-f_{i}^{mes}(\vec{q})\|_{L_{2}(Q)}
\label{error2}
\end{equation}

The solution to the optimization problem is then defined by Equation \ref{min}

\begin{equation}
\vec{p^{*}}=\arg\min_{\vec{p} \in S^{n}}\{e^{FM}(\vec{p})\}
\label{min}
\end{equation}
\noindent
where, $\vec{p^{*}}$ is the best parameters vector and $S^{n}$ is constrained searching domain.

The iterative search algorithm continues until either convergence tolerance, $\varepsilon$, or the maximum number of iterations exceed the specified limit.
The search domain is specified for all $n$ parameters, $S^{n}[\vec{s}^{1} \hspace{3 mm} \vec{s}^{2}]$.
First, the IMCMC algorithm samples randomly and stores $k$ number of parameters sets, $\vec{p^{i}} \in S^{n};i=1,2,...,k$.
Then, the model error functions ,$e^{FM}(\vec{p})$, are calculated and stored in error function vector, $\vec{\mathcal{E}}_{1\times k}=\{e^{fm}(\vec{p^{1}}),e^{fm}(\vec{p^{2}}),...,e^{fm}(\vec{p^{k}})\}$.
Then from the current error function vector, the lowest error value, $\textrm{min}(\vec{\mathcal{E}})=e^{fm}(\vec{p}^{*})$, and the corresponding set of parameters, $\vec{p}^{*}$, are identified.
$\vec{p}^{*}$ is chosen as the closest approximation of the solution to the global minima.

Using IMCMC in complex forward problems with a wide search range generally shows a slow convergence rate. 
Therefore, surrogate models are used to accelerate the entire process by finding local optimum regions.
For this purpose, all the stored sets of parameters and corresponding error sets are used to train the surrogate-model.

We use fully connected deep network \cite{Chow2007} to train the model by all parameter sets and their corresponding error functions.
Equation \ref{NN} presents the mapping system of the trained surrogate-model.

\begin{equation}
\vec{f}^{sm}:(\vec{p})\longrightarrow{e}^{fm}(\vec{p},\vec{q})
\label{NN}
\end{equation}

Once trained, we have a surrogate-model to obtain surrogate model error function, $e^{sm}(\vec{p})$ ,without the computational cost of forward model.
To find the minimum of $e^{sm}(\vec{p})$, we use genetic algorithm (GA).
Using GA, the local minima of the surrogate-model, $\vec{p}^{sm}$, is found per Equation \ref{Addmin}

\begin{equation}
\vec{p}^{sm} = \arg\min_{\vec{p}}\{e^{sm}(\vec{p})\}
\label{Addmin}
\end{equation}

The local minima, $\vec{p}^{sm}$, is added to the sets of the parameter vectors, $\vec{p}$.
Then the model error function, $e^{fm}(\vec{p}^{sm})$, is calculated using the forward model and added to the error function vector, $\vec{\mathcal{E}}$.
If $e^{fm}(\vec{p}^{sm})$ is lower than the convergence tolerance, then the iterations are terminated. Otherwise, the best parameters set, $\vec{p}^{*}$, is updated and the following RS is executed.

New sets of parameter vectors are generated based on normal distribution in the neighborhood of poles, and their corresponding model error functions are evaluated.
Using poles help in the search for the global minimum around all possible local minima.
A total of $j$ number of poles are chosen from the latest parameter vectors, $\vec{p}$ to cover highest probable area.
To choose the highest probable area, all $\vec{p}$ are sorted based on the magnitude of their model error function (lowest to highest error).
Then the poles are selected among the lowest error range.
Since $\vec{p}^{*}$ is the best answer it forms the first pole.
The subsequent poles then are chosen corresponding to the top ranked population (e.g., $10\%$ and $30\%$), $\vec{\mathfrak{P}}=\{\vec{p}^{*},\vec{p}^{2},...,\vec{p}^{j}\}$.

A total of $g$ number of random parameter vectors is generated centered around each pole using a normal distribution.
The generated parameter vectors are added to the entire parameter vectors.
The forward model is then used to estimate the model error function for all newly added parameter vectors and $\vec{p}^{*}$ is updated.
Finally, the search domain is updated around $\vec{p}^{*}$ per Equation \ref{update}

\begin{equation}
S^{n}[\vec{s}^{1} \hspace{3 mm} \vec{s}^{2}] = [a\times\vec{p}^{*} \hspace{3 mm} b\times\vec{p}^{*}]
\label{update}
\end{equation}
where $a$ and $b$ are coefficients adjusting the range.
The surrogate model is trained again and iteration continues until the satisfying the convergence criteria is met or the maximum number of iterations is exceeded.


\section{Result and discussion}
 \label{subsect.RD}

\begin{figure}[ht!]
    \centering
    \subfloat[]{{\includegraphics[width=3.5cm]{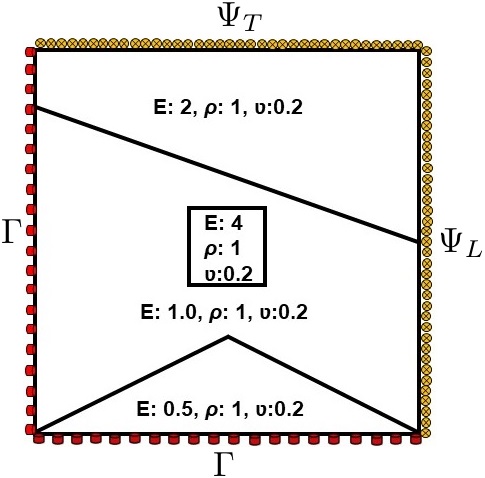} }}%
\quad
    \subfloat[]{{\includegraphics[width=3.5cm]{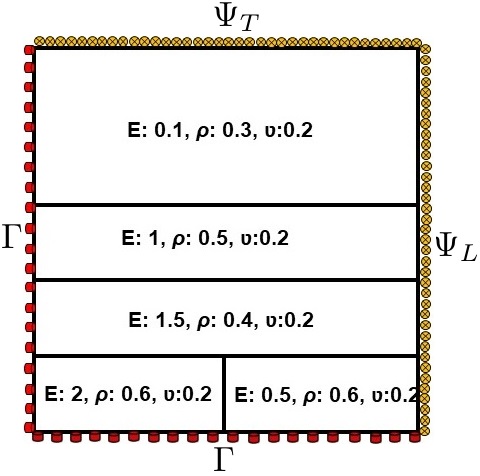} }}%
\quad
    \subfloat[]{{\includegraphics[width=3.5cm]{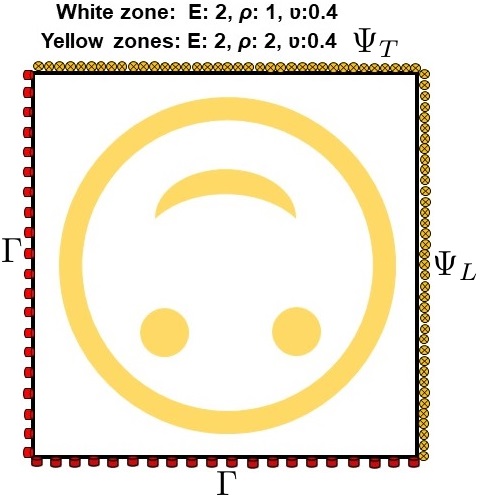} }}%
    \caption{Schematic illustrations of all three simulated experiments.}
    \label{SCHE}
\end{figure}

In this section, we solve three examples to demonstrate the application of proposed method for estimating mechanical properties of solid materials.
All three examples are propagation of stress waves into heterogeneous two dimensional media.
In all examples, the mechanical properties are estimated by sending stress waves from two sides and capturing the strain response at two other sides.
Figure \ref{SCHE} schematically illustrates all three examples.
In all examples, the experimental wave propagations were simulated by solving KT forward model and addition of $0.1\%$ Gaussian noise. 
As measurements, the nodal strains vector is provided over certain boundaries. 
As the boundary conditions, for all examples shown, left and bottom edges, $\Gamma$, velocity controled stress waves are injected and over right and top edges, $\Psi$, nodal strains are recorded as measurements.
The velocity of the excitation wave over $\Gamma$ is defined by Equation \ref{BC1}.
It should be noted that the discretization of the simulated forward models and inverse models were different.

\begin{eqnarray}
v(\eta,t) = sin(2\pi\times t); \hspace{2 mm} \eta \in \Gamma \hspace{25 mm} t \leq 1
\nonumber \\
v(\eta,t) = 0; \hspace{2 mm} \eta \in \Gamma \hspace{40 mm} t > 1
\label{BC1}
\end{eqnarray}

\subsection{Example 1}
\label{subsec.E1}

In the first example, Figure \ref{SCHE}a, we assume the presence of prior knowledge about density and Poisson coefficient equal to 1 $gr/cm^{3}$ and 0.2 respectively.
The objective is to estimate the nodal elastic moduli distributions in domain.
In most tomography problems, the main objective is anomaly and damage detections.
To achieve these objectives, most common approach is to calculate estatic or dynamic elastic moduli distributions over domains.
Example 1, therefore, focuses on reconstruction of elastic moduli distributions over a domain using stress wave propagations.
To do so, the deep learning network should be trained first to connect nodal elastic moduli to nodal strain measurements vector.
Then, using deep learning surrogate IMCMC algorithm, distributions of elastic moduli will be reconstructed.

For this example, the deep learning input layer ($L_{0}$) is a vector of 10000 units using normalized nodal elastic modulus values.
The output layer ($L_{H}$) is a vector of 202 units which uses normalized nodal strain values at $\Psi_{L}$ and $\Psi_{T}$ boundaries as is shown by Figure \ref{SCHE}a.
Total of 30 hidden layers were used in deep learning network in this example where the number of units of these layers started from 10000 ($L_{1}$) and decreased gradually to 300 at the final hidden layer ($L_{H-1}$).
$L_{2}$ regularization parameters were chosen using an initial batch of random parameter vectors as train and development sets to avoid overfitting.
$25 \%$ dropout probability was chosen as the optimum regularization parameter.
The momentum terms, learning rate and learning rate decay all were tuned by training the network based on initial random search batch.
The estimated parameters at each iteration of inverse problem are divided into 10 mini-batches and each set of parameters generated statisticaly around poles are considered new mini-batches.

\begin{figure}[ht!]
    \centering
    \subfloat[True E]{{\includegraphics[width=4cm]{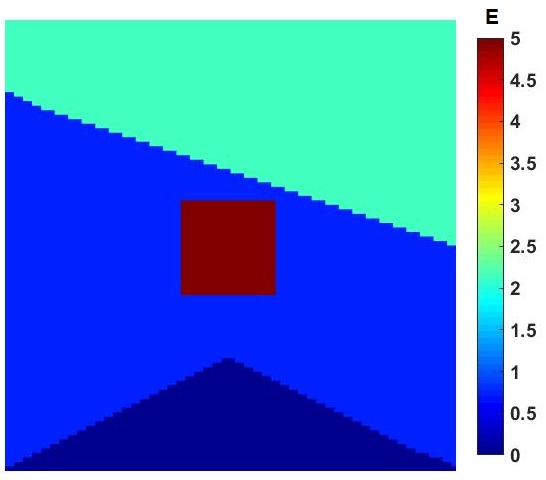} }}%
\quad
	\subfloat[FWI reconstructed E]{{\includegraphics[width=4cm]{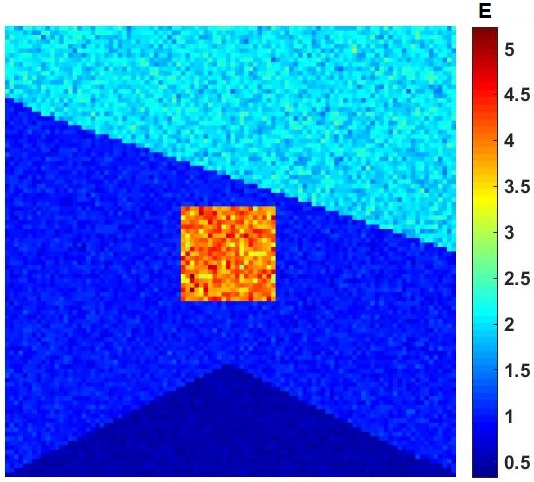} }}%
    \caption{Nodal elastic moduli distribution of the domain shown by Figure \ref{SCHE}a: (a) true elastic moduli, (b) reconstructed elastic moduli.}
    \label{E1_1}
\end{figure}

The deep learning based IMCMC algorithm starts with 100000 initial random parameter vectors batch as random batch.
The search domain was initiated for elastic modulus, $E$ between $[1 \hspace{3 mm} 20 ]$.
The training progress window were $a=0.8$ and $b=1.2$, which means that after surrogate model training, the search domain includes trial solutions within $\pm20\%$ of the optimal solution.
10 statistical search poles were used, $j=10$, around best parameter vector and best $5, 10, 15, 20, 25, 30, 35, 40, 45\%$ parameter vectors.
A total of $g=2000$ random parameter vectors were generated at each pole.
The convergence tolerance for this example was considered $10\%$.
The deep learning search algorithm iterated while the lowest error function is higher than convergence tolerance.
Figure \ref{E1_1}a and \ref{E1_1}b compare the true and the reconstructed elastic modulus distributions.

\begin{figure}[ht!]
    \centering
    \subfloat[]{{\includegraphics[width=5.5cm]{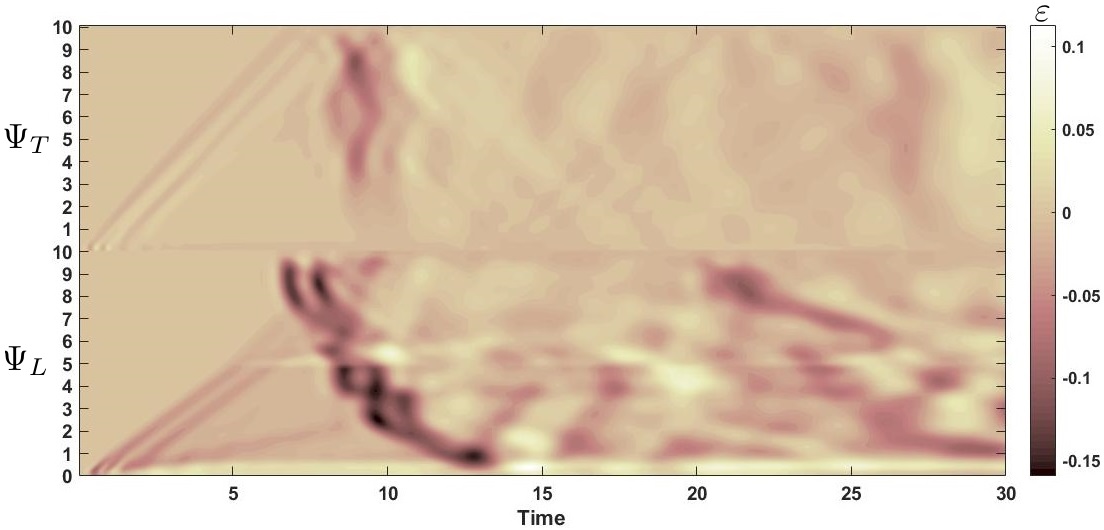} }}%
\quad
	\subfloat[]{{\includegraphics[width=5.5cm]{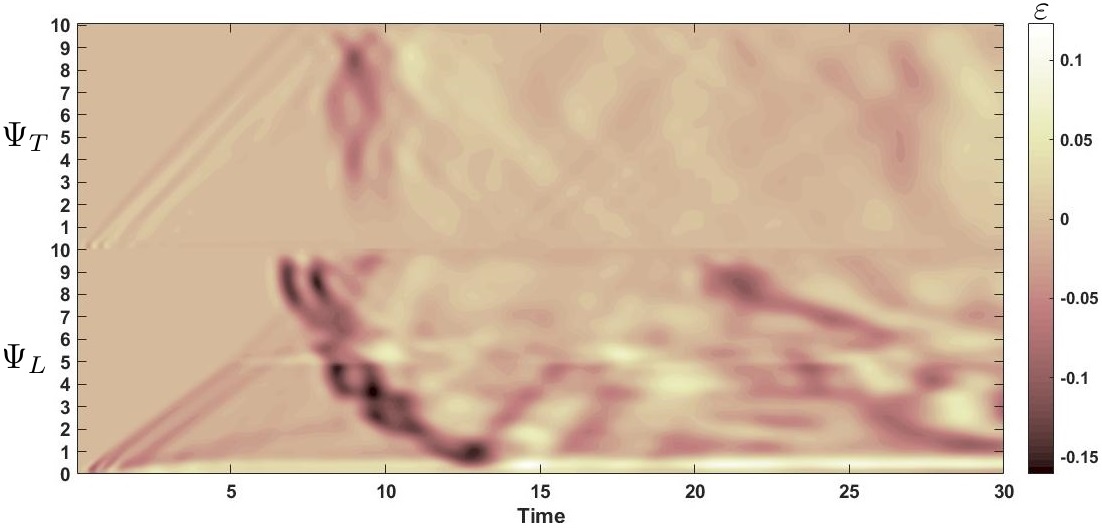} }}%
    \caption{Nodal strain values at $\Psi_{T}$ and $\Psi_{L}$ boundaries with respent to time: (a) true strain measurements, and (b) reconstructed strain values.}
    \label{E1_2}
\end{figure}

The results indicate that the rconstructed elastic moduli distribution is in good agreement with true elastic moduli.
Comparing true and reconstructed results suggest average of $5 \%$ distributed error is measured between true and reconstructed elastic moduli distributions.
Comparing these two images, reconstrcuted image provides proper distinction between different blocks of materials.
Figure \ref{E1_2}a presents measured nodal strain values at $\Psi_{T}$ and $\Psi_{L}$ boundaries over time, which were considered as optimization objectives between real and simulation cases. 
The whole time of $30$ was simulated since it was sufficient to capture almost all propagation events inside the domain for FWI reconstructions.
Figure \ref{E1_2}b presents similar results for recosntructed elastic moduli distribution to \ref{E1_2}a.
The covergence norm between true measurements and estimated one at final iteration was $5.77 \%$ which means true and estimated measurements had average of $\pm 5.77 \%$ difference at each time and space dimensions.

\begin{figure}[ht!]
    \centering
    \subfloat{{\includegraphics[width=2.5cm]{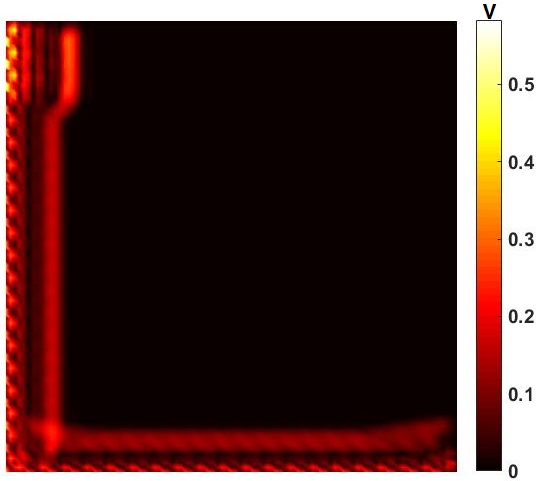} }}%
\smallskip
    \subfloat{{\includegraphics[width=2.5cm]{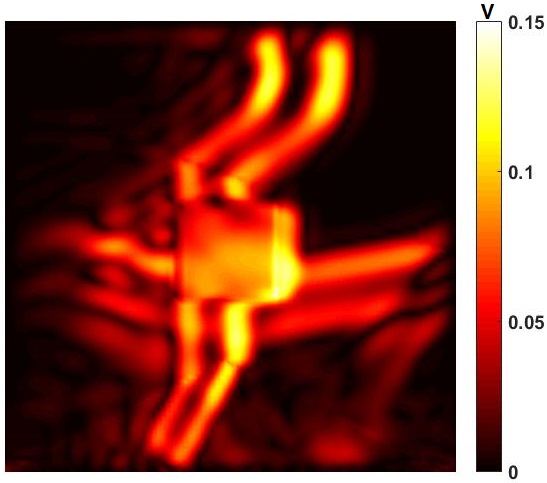} }}%
\smallskip
    \subfloat{{\includegraphics[width=2.5cm]{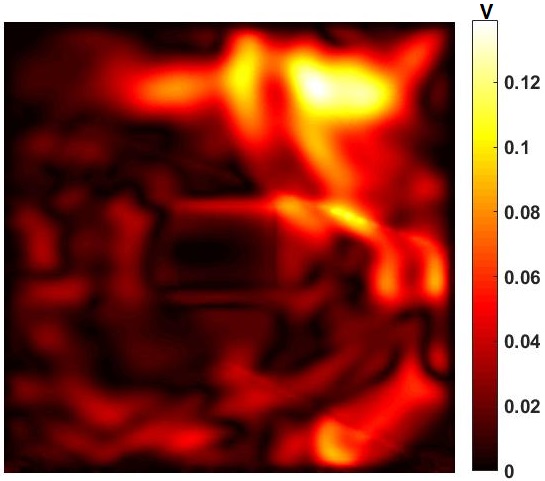} }}%
\smallskip
    \subfloat{{\includegraphics[width=2.5cm]{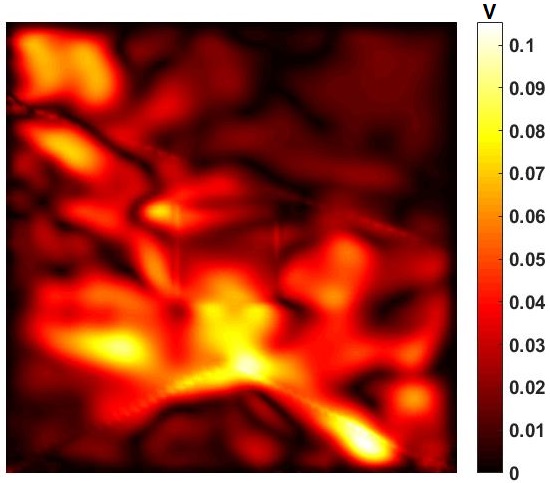} }}%
\smallskip
    \subfloat{{\includegraphics[width=2.5cm]{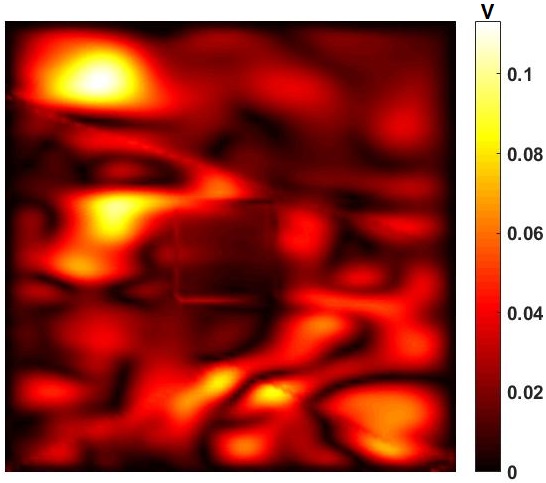} }}%
\\
	\subfloat{{\includegraphics[width=2.5cm]{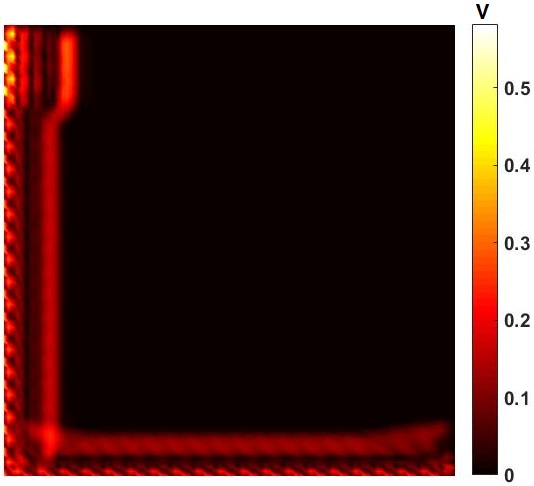} }}%
\smallskip
    \subfloat{{\includegraphics[width=2.5cm]{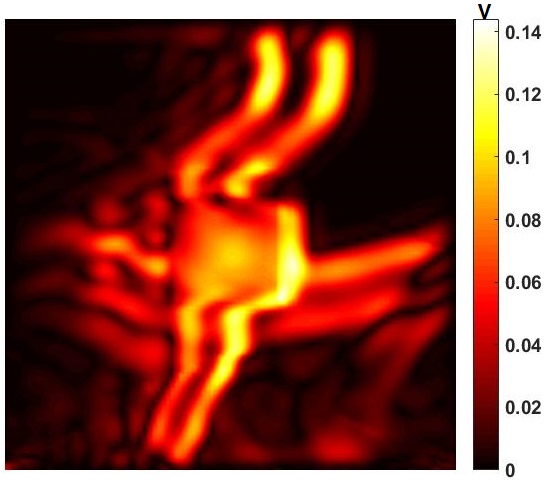} }}%
\smallskip
    \subfloat{{\includegraphics[width=2.5cm]{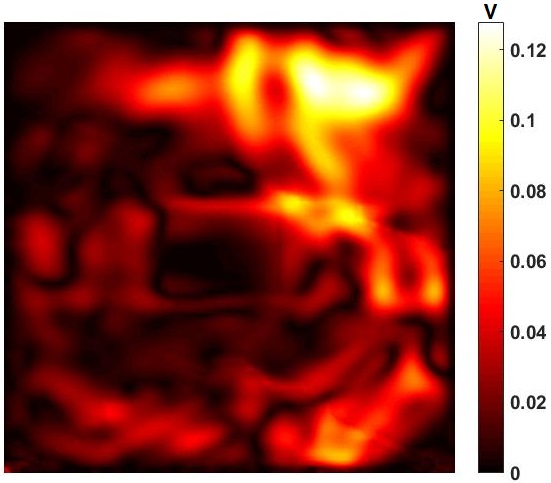} }}%
\smallskip
    \subfloat{{\includegraphics[width=2.5cm]{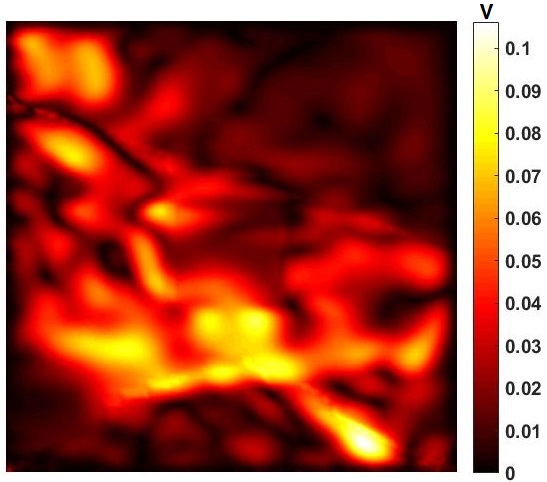} }}%
\smallskip
    \subfloat{{\includegraphics[width=2.5cm]{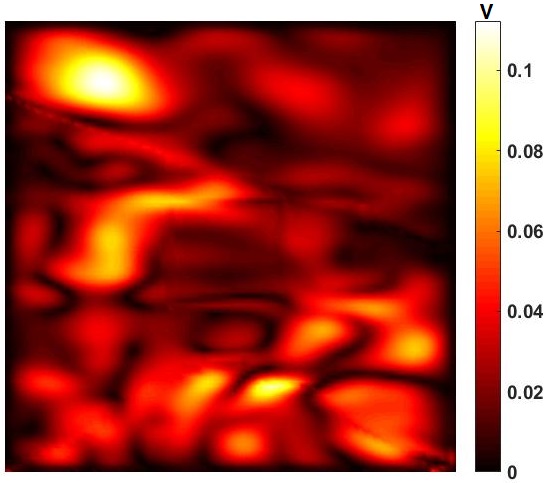} }}%
    \caption{Wave velocity distribution inside domain: (top row) inside real domain at 1, 5, 10, 15 and 30 seconds respectively from left to right, (bottom row)  inside FWI reconstructed domain at 1, 5, 10, 15 and 30 seconds respectively from left to right.}
    \label{E1_3}
\end{figure}

Finally, Figure \ref{E1_3} presents comparison of wave velocity distributions between real and recosntrcuted domain at 1, 5, 10, 15 and 30 seconds.
The top row of Figure \ref{E1_3} presents wave velocity distribution inside real domain at 1, 5, 10, 15 and 30 seconds respectively from left to right.
The bottom row of Figure \ref{E1_3} presents wave velocity distribution inside reconstructed domain at 1, 5, 10, 15 and 30 seconds respectively from left to right.
Similarly, Figure \ref{E1_3} suggests very good agreement between both situations.

\subsection{Example 2}
\label{subsec.E2}

In the second example, Figure \ref{SCHE}b, we only assume prior information about Poisson coefficient equal to 0.2.
Therefore, the objective is to estimate distributions of elastic moduli and densities over the domain.
Often, the objective of tomography problems exceeds to estimate real material properties.
To achieve such objective, most common approach is to estimate wave velocity distributions over domains which contain effects of elastic modulus and density.
Using common approaches of wave travel time, gradient based inverse methods, and generally qualitative tomographic methods, it is a very cumbersome objective to estimate elastic moduli and densities distribution at same time seperately.
Also, due to probable high ill-posedness of inverse problem, it is not guaranteed to estimate the most optimum responses.
To achieve this objective, here, the deep learning network was trained first to connect vector of nodal elastic moduli and densities to nodal strain measurements vector.
Then, using the optimization of deep learning model, distributions of elastic moduli and densities are estimated at same time.

In this example, the deep learning input layer ($L_{0}$) was a vector of 20000 units uses normalized nodal elastic moduli and densities at same time.
The output layer ($L_{H}$) is a vector of 202 units which uses normalized nodal strain values at $\Psi_{L}$ and $\Psi_{T}$ boundaries as is shown by Figure \ref{SCHE}b.
Total of 45 hidden layers were used in deep learning network in this example where the number of units of these layers started from 20000 ($L_{1}$) and decreased gradually to 202 at the final hidden layer ($L_{H-1}$).
Similarly, $L_{2}$ regularization and dropout method were used.
$35 \%$ dropout probability was chosen as the optimum regularization parameter.
The dropout probability here was chosen higher than first example to introduce higher randomness and avoid vanishing issue in this example.
Similar momentum terms, learning rate and learning rate decay were used in this example as example 1.
After each inverse problem iteration, new parameters are divided into 10 mini-batches and each set of parameters generated statisticaly around poles are considered new mini-batches.

\begin{figure}[ht!]
    \centering
    \subfloat[True E]{{\includegraphics[width=4cm]{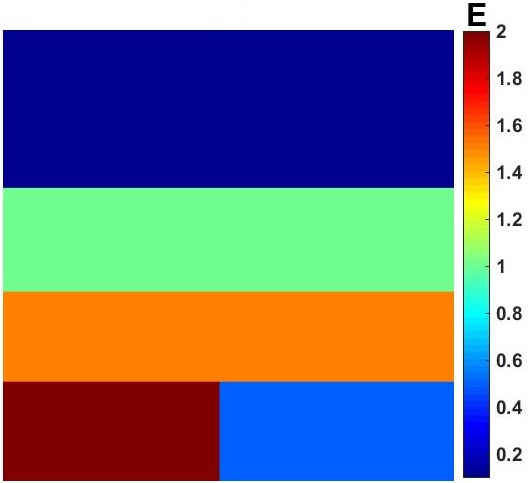} }}%
\quad
	\subfloat[FWI reconstructed E]{{\includegraphics[width=4cm]{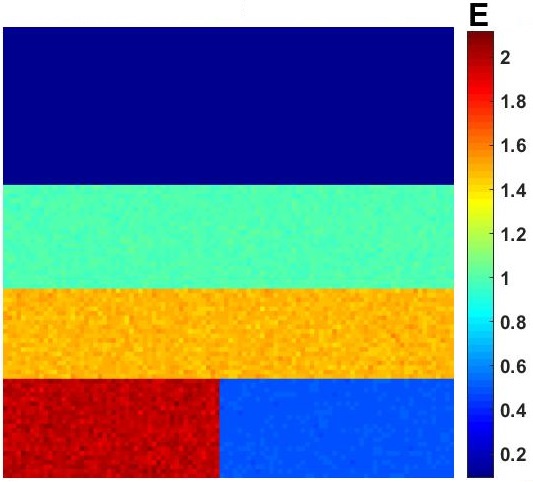} }}%
\\
    \subfloat[True $\rho$]{{\includegraphics[width=4cm]{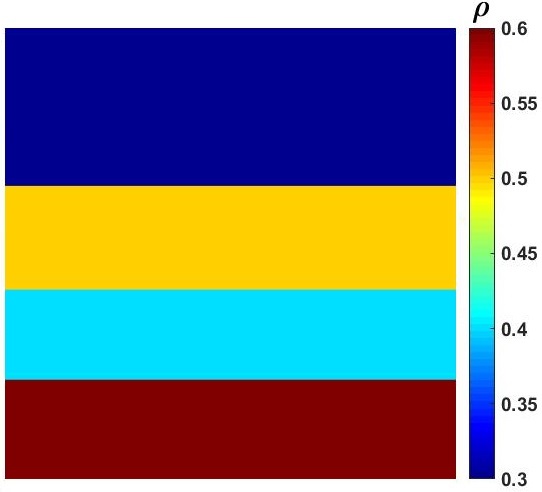} }}%
\quad
	\subfloat[FWI reconstructed $\rho$]{{\includegraphics[width=4cm]{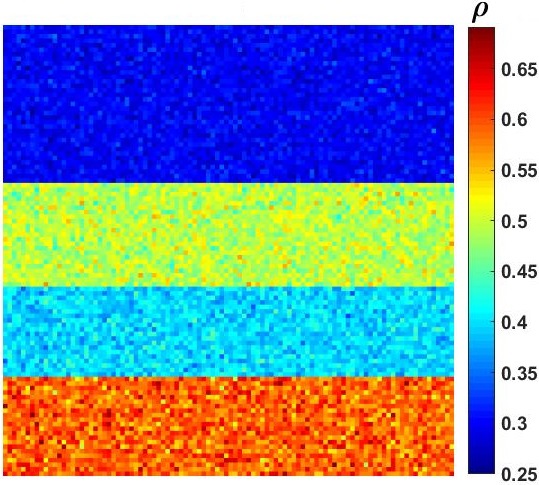} }}%
    \caption{(a) True elastic moduli distribution, (b) FWI reconstructed elastic moduli distribution,  (c) True densities distribution, (d) FWI reconstructed densities distribution.}
    \label{E2_1}
\end{figure}

The deep learning based IMCMC starts with 300000 initial random parameter vectors batch as random batch.
The search domain was initiated for elastic modulus, $E$ between $[1 \hspace{3 mm} 20 ]$; and for density , $\rho$ between $[0.001 \hspace{3 mm} 2 ]$ (used higher than zero to avoid singularity).
The training window were $a=0.8$ and $b=1.2$ to search parameters within $\pm20\%$ of the optimal solution in every iteration.
10 statistical search poles were used, $j=10$, around best parameter vector and best $5, 10, 15, 20, 25, 30, 35, 40, 45\%$ parameter vectors.
A total of $g=4000$ random parameter vectors were generated at each poles.
The convergence tolerance for this example was considered $10\%$, however, the inverse problem had not converged to less than $10\%$ before the iteration number exceeds iteration limit.
Comparing true and reconstructed material properties, Figure \ref{E2_1} compares the true and the reconstructed elastic moduli and dnsities distributions.

According to Figure \ref{E2_1}, the results indicate that the reconstructed elastic moduli and densities distributions are in agreement with true ones.
Comparing true and reconstructed $E$ results suggest average of $1.3 \%$ distributed error is measured between true and reconstructed elastic modulus distributions.
Comparing true and reconstructed $\rho$ results suggest average of $1 \%$ distributed error is measured between true and reconstructed density distributions.
Comparing these two columns, proposed FWI deep learning method succeeded to reconstruct both $E$ and $\rho$ distributions simultaneously.

\begin{figure}[ht!]
    \centering
    \subfloat{{\includegraphics[width=5.5cm]{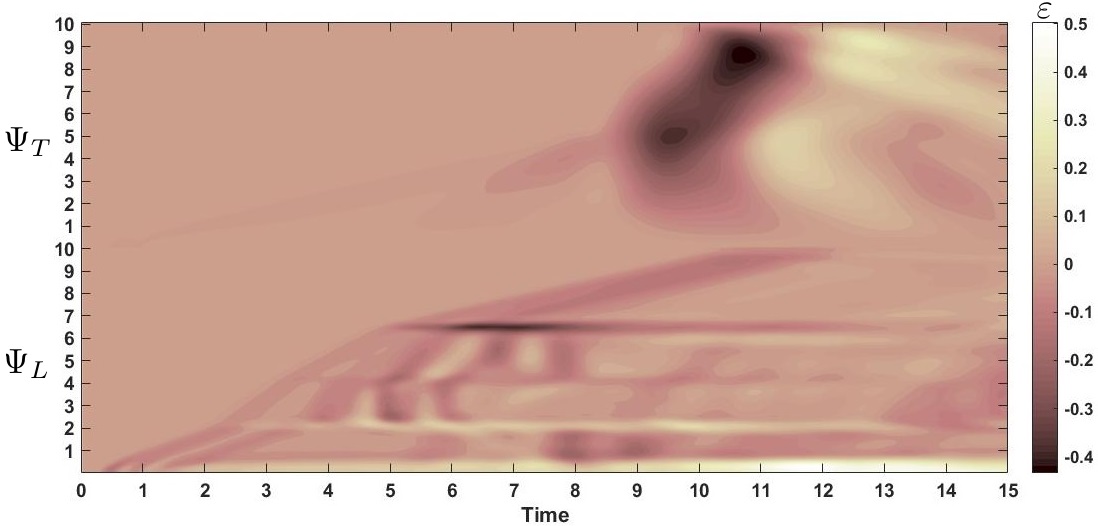} }}%
\quad
	\subfloat{{\includegraphics[width=5.5cm]{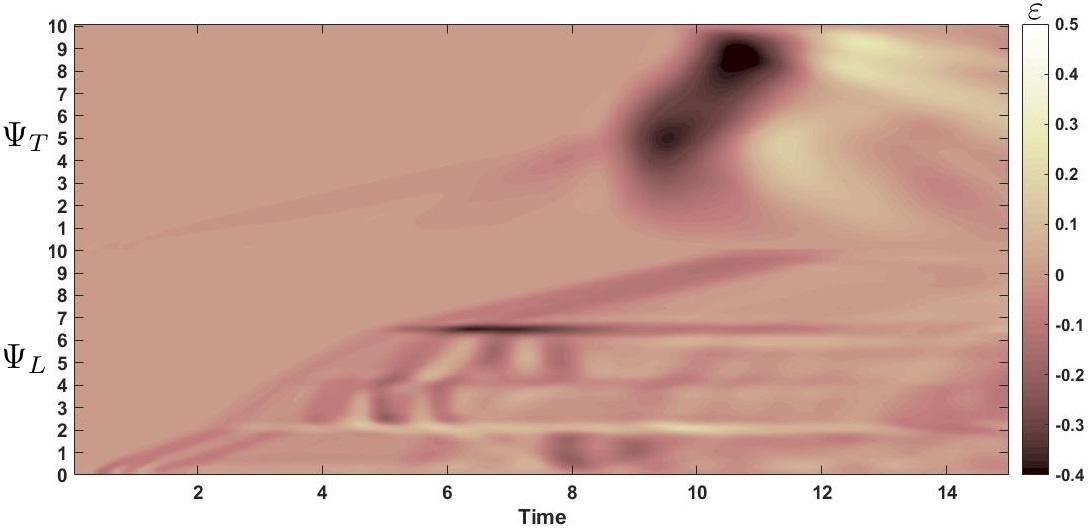} }}%
    \caption{Nodal strain values at $\Psi_{T}$ and $\Psi_{L}$ boundaries with respent to time: (a) true strain measurements, and (b) FWI reconstructed measurements.}
    \label{E2_2}
\end{figure}

Figure \ref{E2_2} presents measured nodal strain values over $\Psi_{T}$ and $\Psi_{L}$ at whole 15 time range which were considered as optimization objectives between real and simulation cases.
The covergence norm between true measurements and estimated one at final iteration was $14.99 \%$ which means variation of estimated measurement is $\pm 14.99 \%$ to true measurements.
In this example, the covergence condition of $10 \%$ has not been met.
Therefore, after iterations exceeded iteration limit, the inverse problem was terminated.
Reconstruction of both densities and elastic moduli at the same time introduces higher ununiqueness and non-convexity to the inverse problem.
That is why the inverse problem was not able to converge to the $10 \%$ $L-{2}$ norm convergence.
Comparing two results of Figure \ref{E2_2}, measurements are still comparable.

\begin{figure}[ht!]
    \centering
    \subfloat{{\includegraphics[width=3cm]{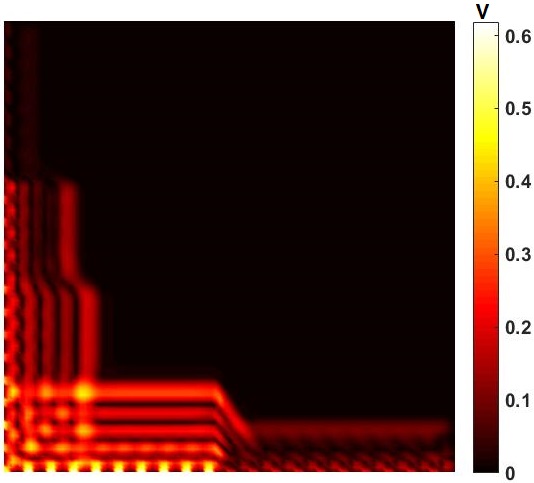} }}%
\smallskip
    \subfloat{{\includegraphics[width=3cm]{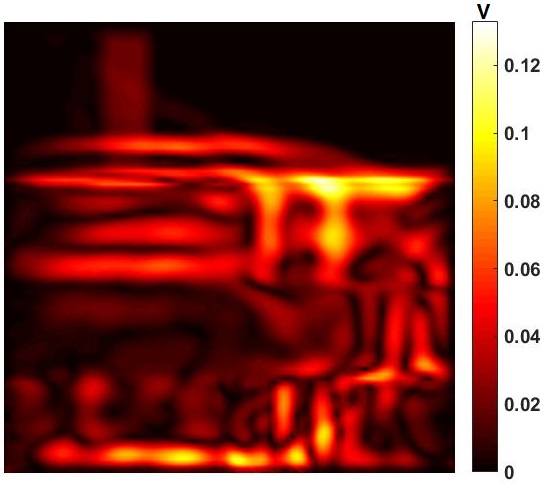} }}%
\smallskip
    \subfloat{{\includegraphics[width=3cm]{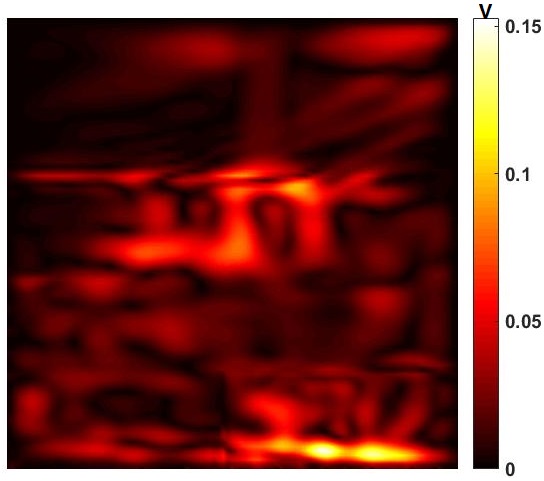} }}%
\smallskip
    \subfloat{{\includegraphics[width=3cm]{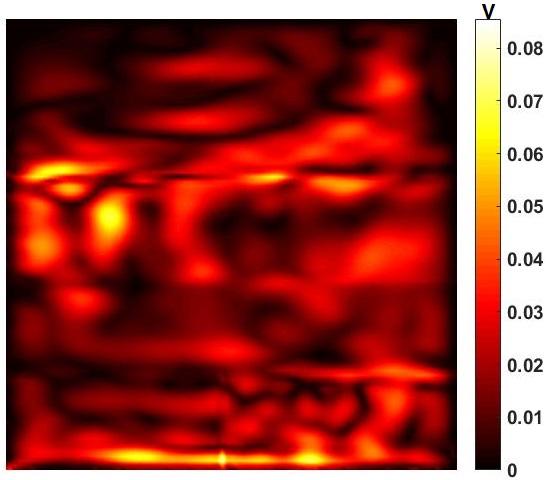} }}%
\\
	\subfloat{{\includegraphics[width=3cm]{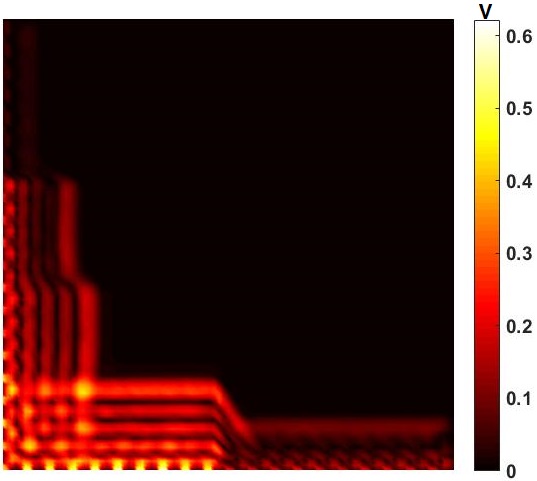} }}%
\smallskip
    \subfloat{{\includegraphics[width=3cm]{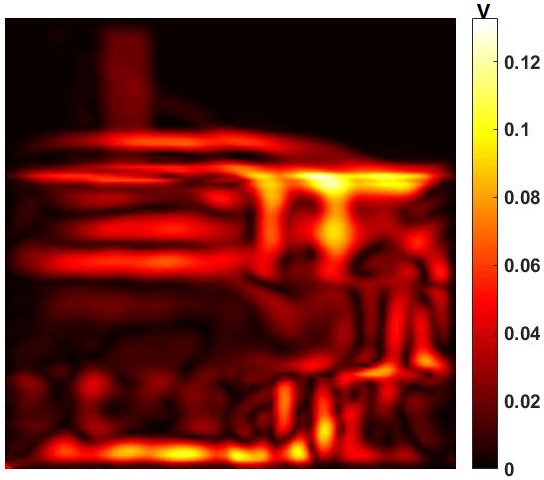} }}%
\smallskip
    \subfloat{{\includegraphics[width=3cm]{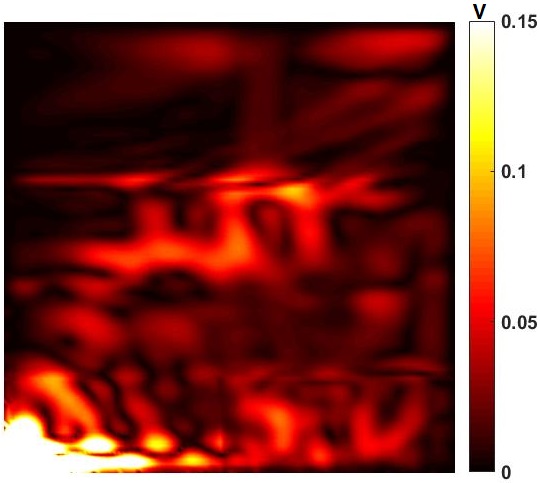} }}%
\smallskip
    \subfloat{{\includegraphics[width=3cm]{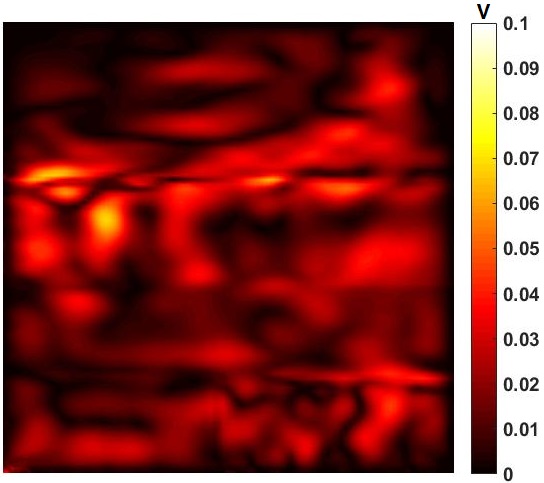} }}%
    \caption{Wave velocity distribution inside domain: (top row) inside real domain at 1, 5, 10 and 15 seconds respectively from left to right, (bottom row)  inside FWI reconstructed domain at 1, 5, 10 and 15 seconds respectively from left to right.}
    \label{E2_3}
\end{figure}

Finally, Figure \ref{E2_3} presents comparison of wave velocity distributions between real and recosntrcuted domain at 1, 5, 10 and 15 seconds.
The top row of Figure \ref{E2_3} presents wave velocity distribution inside real domain at 1, 5, 10 and 15 seconds respectively from left to right.
The bottom row of Figure \ref{E2_3} presents wave velocity distribution inside reconstructed domain at 1, 5, 10 and 15 seconds respectively from left to right.
Reconstructing both densities and elastic moduli at the same time using transient wave propagation can be cumbersome for main two reasons.
First, simulation of wave propagation through realistic detailed volumetric representations of heterogeneous materials is cumbersome because of the huge computational requirements for higher order simulation methods and high chance of unstability of lower order methods.
Numerical methods to accurately simulate the wave equation are being constantly developed and applied with increasing levels of sophistication. 
KT, has been shown to be an efficient method for simulation of nonlinear conservation differential equations with high resolution and can handdle high scattering and shock waves.
Second, variation of density and elastic modulus together can raise huge amount of non-convexity.
Here, comparing results, shows that using proposed deep learning based inverse model and KT forward model have been able to handle these complexities.

\subsection{Example 3}
\label{subsec.E3}

In the third example, Figure \ref{SCHE}c, we consider Poisson coefficient equal to 0.2 and known nodal elastic moduli, therefore, the objective is to estimate densities distribution of the domain.
This can be helpful for cases that different materials with different densities have similar elastic moduli.
For this example, similar model of deep learning to example 1 was used.
However, wave propagation simulation is more sensitive to variation of densities than variation of elastic moduli.
Also, variation of densities cause higher ill-posedness of inverse problem.
Therefore, here, the deeper learning network was trained to connect vector of nodal densities to the nodal strain measurements.

Similar to example 1, the deep learning input layer ($L_{0}$) was a vector of 10000 units uses normalized nodal elastic moduli and densities at same time.
The output layer ($L_{H}$) is a vector of 202 units which uses normalized nodal strain values at $\Psi_{L}$ and $\Psi_{T}$ boundaries as is shown by Figure \ref{SCHE}b.
Total of 30 hidden layers were used in deep learning network in this example where the number of units of these layers started from 10000 (($L_{1}$)) and decreased gradually to 202 at the final hidden layer ($L_{H-1}$).
Similarly, $L_{2}$ regularization and dropout method were used.
$30 \%$ dropout probability was chosen as the optimum regularization parameter.
The dropout probability here was chosen higher than first example to introduce higher randomness in this example.
Similar momentum terms, learning rate and learning rate decay were used in this example as example 1.
The RS based parameters are divided into 10 mini-batches and each set of parameters generated statisticaly around poles are considered new mini-batches.

\begin{figure}[ht!]
    \centering
    \subfloat[True $\rho$]{{\includegraphics[width=4cm]{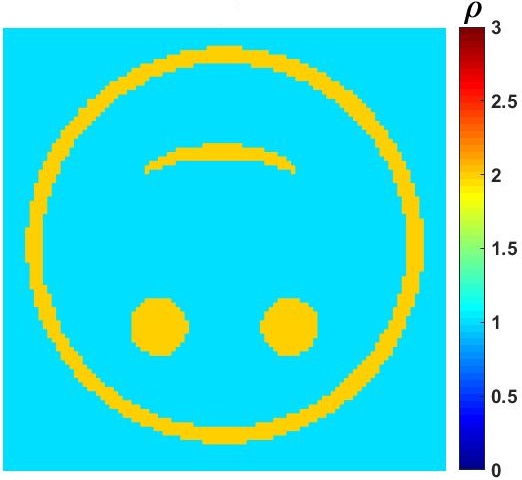} }}%
\quad
	\subfloat[FWI reconstructed $\rho$]{{\includegraphics[width=4cm]{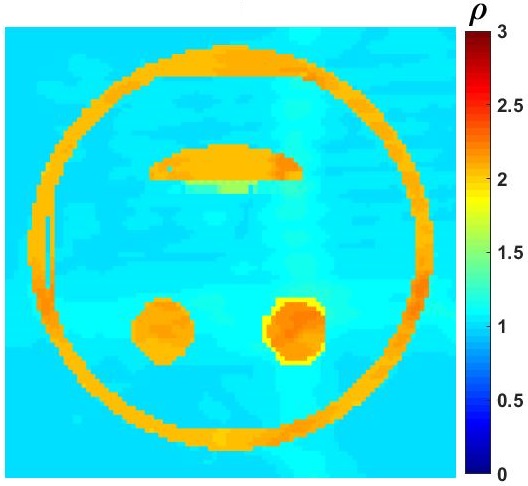} }}%
    \caption{(a) True densities distribution, (b) FWI reconstructed densities distribution.}
    \label{E3_1}
\end{figure}

The deep learning based ICMC algorithm starts with 100000 initial random parameter vectors batch as random batch.
The search domain was initiated for density , $\rho$ between $[0.001 \hspace{3 mm} 5 ]$ (used higher than zero to avoid singularity).
The training progress window were $a=0.8$ and $b=1.2$ to search RS parameters within $\pm20\%$ of the optimal solution in every iteration.
10 statistical search poles were used, $j=10$, around best parameter vector and best $5, 10, 15, 20, 25, 30, 35, 40, 45\%$ parameter vectors.
A total of $g=4000$ random parameter vectors were generated at each poles.
Comparing true and reconstructed material properties, Figure \ref{E3_1} compares the true and the reconstructed densities distributions.

According to Figure \ref{E3_1}, the results indicate that the reconstructed densities distributions are in agreement with true ones.
Comparing true and reconstructed $\rho$ results suggest average of $3.2 \%$ distributed error is measured between true and reconstructed elastic modulus distributions.
Comparing these results, proposed method succeeded to reconstruct $\rho$ distributions in good agreement to the real case.

\begin{figure}[ht!]
    \centering
    \subfloat{{\includegraphics[width=5.5cm]{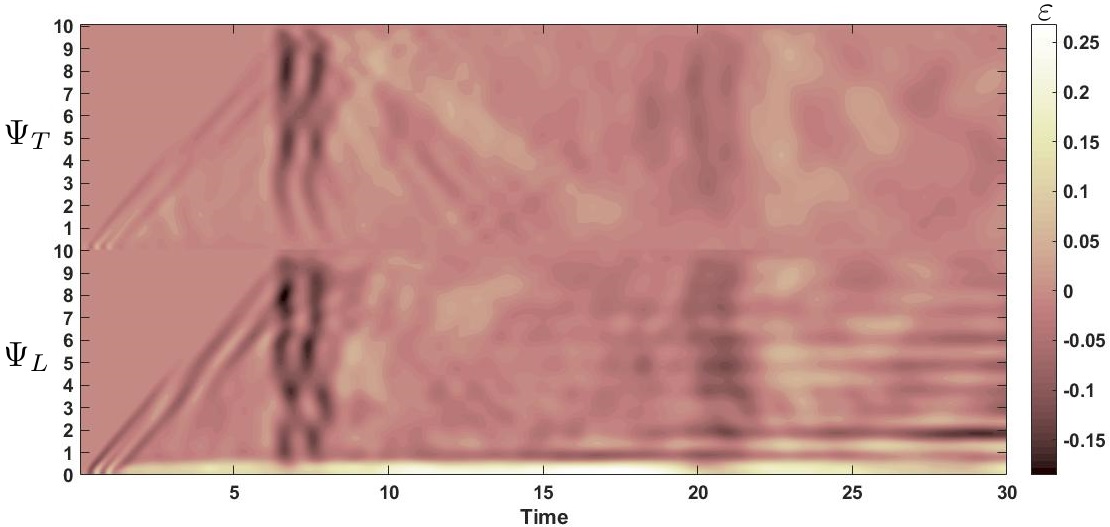} }}%
\quad
	\subfloat{{\includegraphics[width=5.5cm]{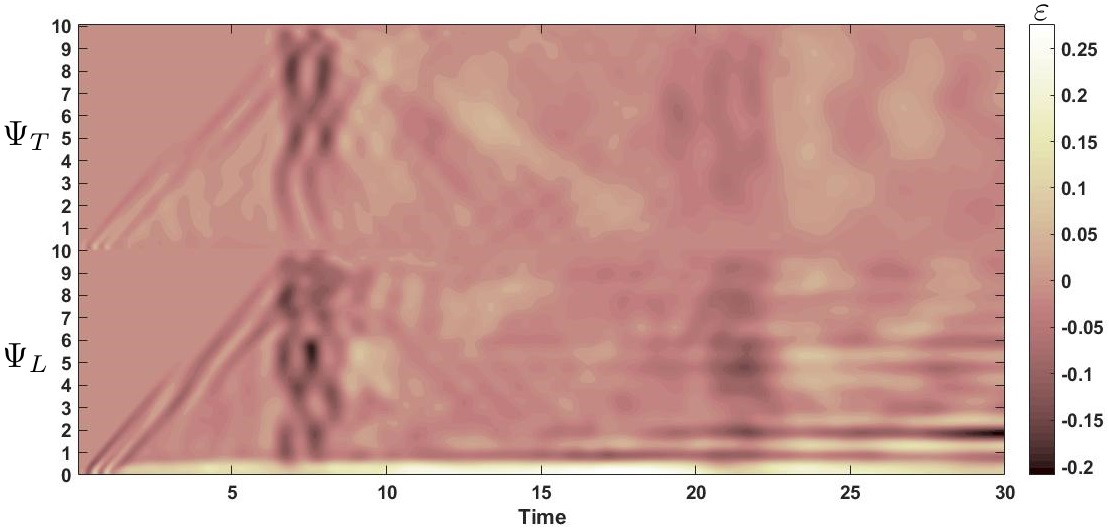} }}%
    \caption{Nodal strain values at $\Psi_{T}$ and $\Psi_{L}$ boundaries with respent to time: (a) true strain measurements, and (b) FWI reconstructed measurements.}
    \label{E3_2}
\end{figure}

Figure \ref{E3_2} presents measured nodal strain values over $\Psi_{T}$ and $\Psi_{L}$ at whole 30 time range which were considered as optimization objectives between real and simulation cases.
The covergence norm between true measurements and estimated one at final iteration was $8.7 \%$ which means variation of estimated measurement is $\pm 8.7 \%$ to true measurements.

\begin{figure}[ht!]
    \centering
    \subfloat{{\includegraphics[width=2.5cm]{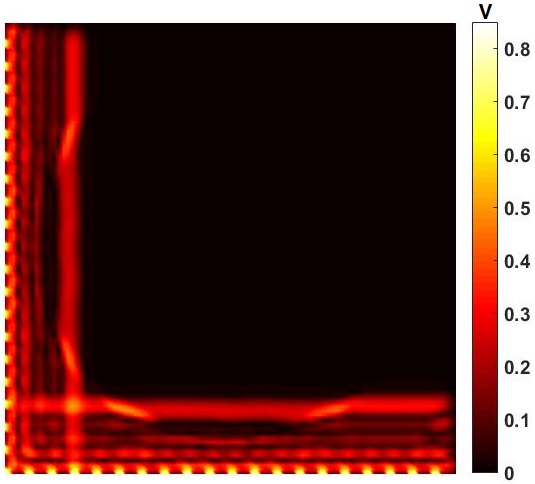} }}%
\smallskip
    \subfloat{{\includegraphics[width=2.5cm]{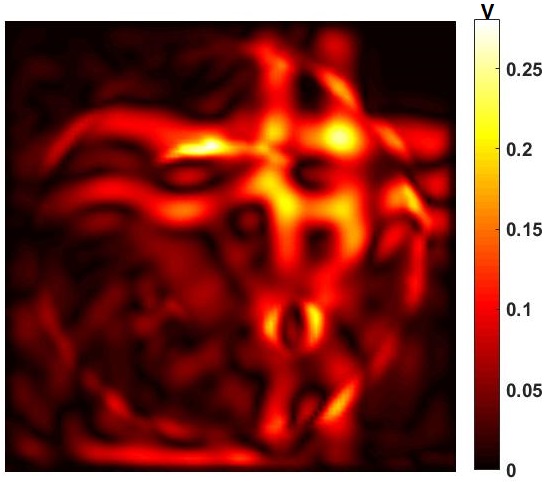} }}%
\smallskip
    \subfloat{{\includegraphics[width=2.5cm]{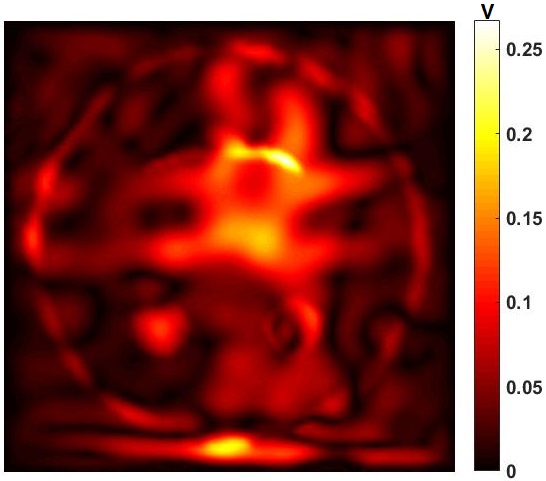} }}%
\smallskip
    \subfloat{{\includegraphics[width=2.5cm]{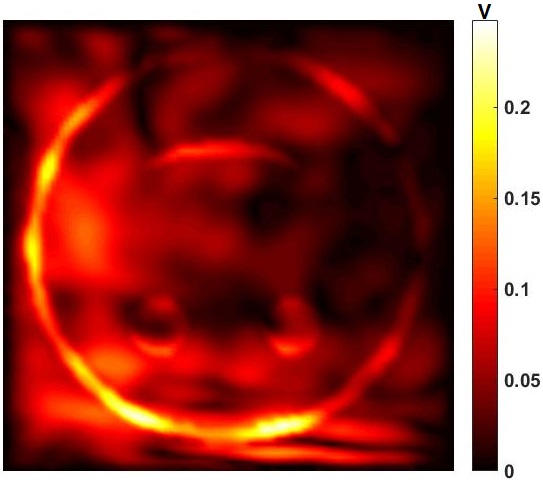} }}%
\smallskip
    \subfloat{{\includegraphics[width=2.5cm]{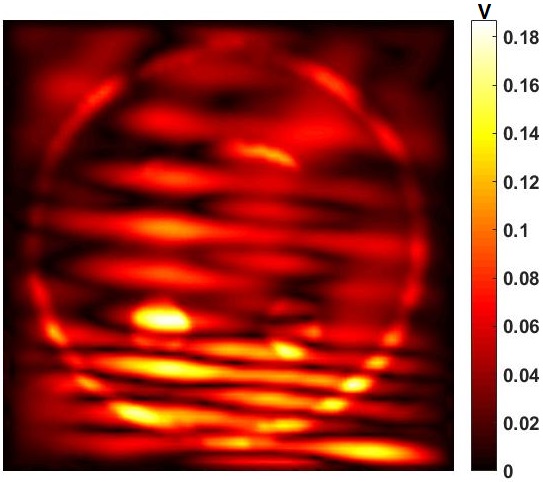} }}%
\\
	\subfloat{{\includegraphics[width=2.5cm]{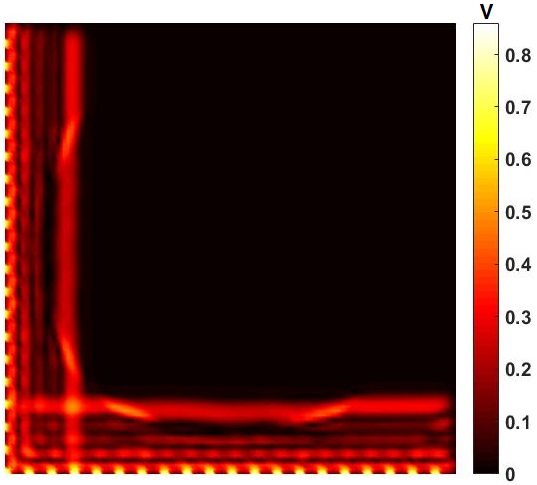} }}%
\smallskip
    \subfloat{{\includegraphics[width=2.5cm]{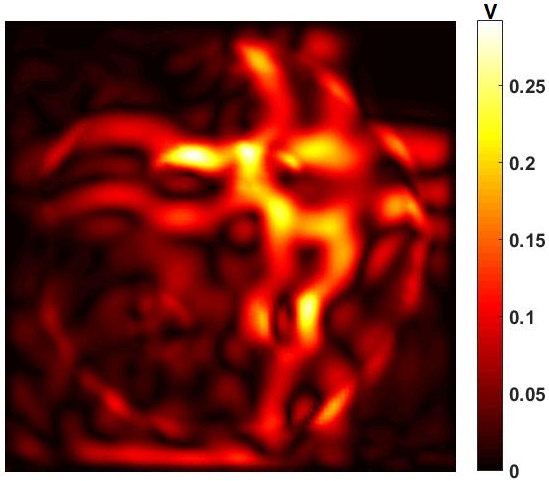} }}%
\smallskip
    \subfloat{{\includegraphics[width=2.5cm]{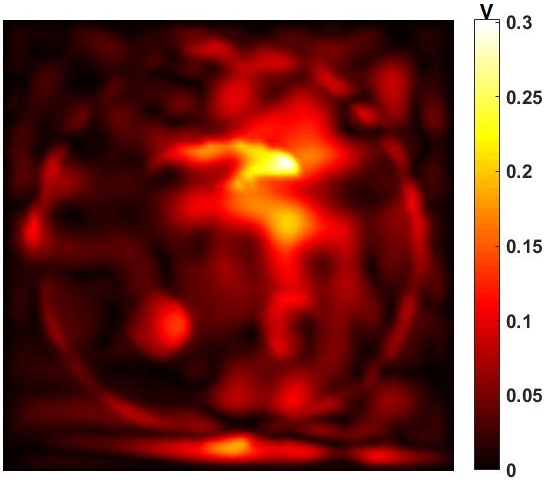} }}%
\smallskip
    \subfloat{{\includegraphics[width=2.5cm]{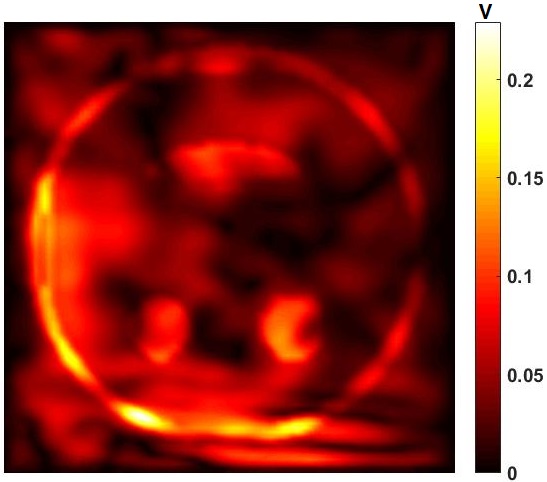} }}%
\smallskip
    \subfloat{{\includegraphics[width=2.5cm]{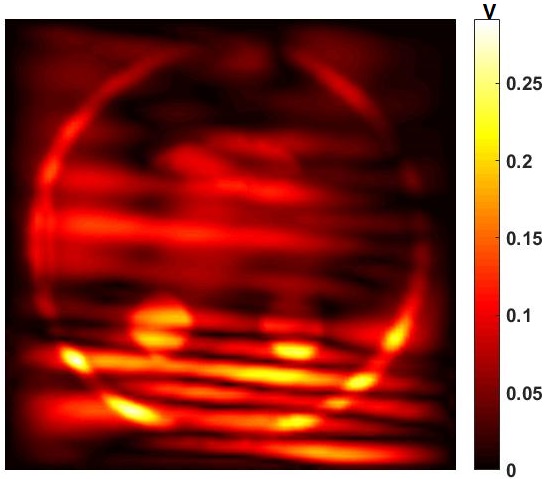} }}%
    \caption{Wave velocity distribution inside domain: (top row) inside real domain at 1, 5, 10, 15 and 30 seconds respectively from left to right, (bottom row)  inside FWI reconstructed domain at 1, 5, 10, 15 and 30 seconds respectively from left to right.}
    \label{E3_3}
\end{figure}

Finally, Figure \ref{E3_3} presents comparison of wave velocity distributions between real and recosntrcuted domain at 1, 5, 10, 15 and 30 seconds.
The top row of Figure \ref{E3_3} presents wave velocity distribution inside real domain at 1, 5, 10, 15 and 30 seconds respectively from left to right.
The bottom row of Figure \ref{E3_3} presents wave velocity distribution inside reconstructed domain at 1, 5, 10, 15 and 30 seconds respectively from left to right.
Reconstructing both densities and elastic moduli at the same time using transient wave propagation can be cumbersome for main two reasons.
First, simulation of wave propagation through realistic detailed volumetric representations of heterogeneous materials is cumbersome because of the huge computational requirements for higher order simulation methods and high chance of unstability of lower order methods.
Numerical methods to accurately simulate the wave equation are being constantly developed and applied with increasing levels of sophistication. 
Kurganov-Tadmor, has been shown to be an efficient method for simulation of nonlinear conservation differential equations with high resolution and can handdle high scattering and shock waves.
Second, vriation of density and elastic modulus together can raise huge amount of non-convexity.
Here, comparing results, shows that using proposed deep learning based inverse model and KT forward model have been able to handle these complexities.

\section{Conclusions}
\label{sect.Conclusions}

We proposed a deep learning surrogate IMCMC algorithm for quantitatively estimating the mechanical properties in heterogeneous media.
In this inverse problem, we utilized high resolution central scheme, Kurganov-Tadmor (KT), as a forward model for stress wave propagation.
This forward model well handles material heterogeneity and gradients.
While this inverse problem is ill-posed and non-convex, this algorithm is successfully used to estimate the mechanical properties.
This appraoch is used instead of end to end deep learning model for connecting nodal strain measurements to nodal elastic moduli due to high ill-posedness and lower accuracy of reconstructions.
Comparing results of reconstructions to real values suggest the deep learning surrogate IMCMC method is able to quantify material characteristics properly.

\bibliography{arxiv}
\bibliographystyle{elsarticle-num}

\end{document}